\documentclass[12pt]{iopart}
\usepackage{iopams}

\expandafter\let\csname equation*\endcsname=\relax
\expandafter\let\csname endequation*\endcsname=\relax
\usepackage{amsfonts,amssymb,amsmath,mathtools,cite,slashed}

\usepackage{graphicx}
\usepackage[all]{xy}
\usepackage{lmodern}
\usepackage[french,english]{babel}

\usepackage{tikz}
\usetikzlibrary{shapes.misc,arrows,decorations.markings}

\DeclareMathOperator{\Dom}{Dom}      
\DeclareMathOperator{\Id}{Id}                 
\DeclareMathOperator{\Ker}{Ker}           
\DeclareMathOperator{\Res}{Res}         
\DeclareMathOperator{\Vol}{\text{Vol}}   

\newtheorem{assumption}{Assumption}[section]
\newtheorem{theorem}[assumption]{Theorem}

\newtheorem{corollary}[assumption]{Corollary}

\newtheorem{definition}[assumption]{Definition}
\newtheorem{prop}[assumption]{Proposition}

\newcommand{\A}{\mathcal{A}}               
\renewcommand{\a}{\alpha}                     
\newcommand{\C}{\mathbb{C}}               
\newcommand{\DD}{\mathcal{D}}            
\newcommand{\DDD}{\vert \mathcal{D} \vert}  
\newcommand{\Ds}{{D\mkern-11.5mu/\,}} 
\newcommand{\Dslash}{{D\mkern-11.5mu/\,}} 
\newcommand{\Ga}{\Gamma}                   
\newcommand{\ga}{\gamma}                    
\renewcommand{\H}{\mathcal{H}}            
\newcommand{\la}{\lambda}                      
\newcommand{\<}{\langle}

\newcommand{\N}{\mathbb{N}}               
\newcommand{\R}{\mathbb{R}}                

\newcommand{\set}[1]{\{\,#1\,\}}               
\renewcommand{\SS}{\mathcal{S}}         
\newcommand{\T}{\mathbb{T}}                 
\newcommand{\wt}{\widetilde}                  
\newcommand{\Z}{\mathbb{Z}}                  
\newcommand{\vc}{\vcentcolon =}             

\def\ee_#1{e_{{\scriptscriptstyle#1}}}       
\def\<#1,#2>{\langle#1\,,\,#2\rangle}          
\newcommand{\norm}[1]{\left\lVert#1\right\rVert}    

\newcommand{\be}{\begin{enumerate}}

\def\Xint#1{\mathchoice
	{\XXint\displaystyle\textstyle{#1}}%
	{\XXint\textstyle\scriptstyle{#1}}%
	{\XXint\scriptstyle\scriptscriptstyle{#1}}%
	{\XXint\scriptscriptstyle\scriptscriptstyle{#1}}%
	\!\int}
\def\XXint#1#2#3{{\setbox0=\hbox{$#1{#2#3}{\int}$}
	\vcenter{\hbox{$#2#3$}}\kern-0.5\wd0}}
\def\ncint{\Xint-}

\begin{document}

\title[]{Global and local aspects of spectral actions}

\author{B. Iochum$^1$, C. Levy$^2$ and D. V. Vassilevich$^{3,4}$}

\address{$^1$ Aix-Marseille Universit\'e and CNRS, Centre de Physique Th\' eorique, Luminy, 
13288 Marseille, France}
\address{$^2$ University of Potsdam, Am Neuen Palais 10, Potsdam, Germany}
\address{$^3$ CMCC - Universidade Federal do ABC, Santo Andr\'e, SP, Brazil}
\address{$^4$ Department of Theoretical Physics, St.Petersburg State University,
St.Petersburg, Russia}
\eads{\mailto{bruno.iochum@cpt.univ-mrs.fr}, \mailto{levy@math.uni-potsdam.de}, \mailto{dvassil@gmail.com}}
\begin{abstract}
The principal object in noncommutatve geometry is the spectral triple consisting of an algebra $\A$, a Hilbert 
space $\H$, and a Dirac operator $\DD$. Field theories are incorporated in this approach by the spectral action
principle, that sets the field theory action to ${\rm Tr}\,f(\DD^2/\Lambda^2)$, where $f$ is a real function such that 
the trace exists, and $\Lambda$ is a cutoff scale. In the low-energy (weak-field) limit the spectral
action reproduces reasonably well the known physics including the standard model.
However, not much is known about the spectral action beyond the low-energy approximation.
In this paper, after an extensive introduction to spectral triples and spectral actions, we study various expansions 
of the spectral actions (exemplified by the heat kernel). We derive the convergence criteria. For a commutative 
spectral triple, we compute the heat kernel on the torus up the second order in gauge connection
and consider limiting cases.
\end{abstract}

\maketitle

\section{Introduction}
Spectral functions of (pseudo)differential operators are being used in mathematical
physics for a long time. For example, the heat kernel was introduced in the context
of quantum physics by Fock \cite{Fock:1937dy} already in 1930's. The zeta regularization,
first suggested by Dowker and Critchley \cite{Dowker:1975tf} and then developed
by Hawking \cite{Hawking:1976ja}, is now the most advanced regularization technique
for a quantum field theory on curved or otherwise complicated backgrounds.

It is widely known that geometry of a manifold is intimately connected to the spectrum
of natural differential operators. A proposal by Connes goes further \cite{Book,Polaris}.
In his approach, geometry of $M$ is defined by a differential operator, or, more precisely,
by a triple consisting of an algebra $\A$ that plays the role of an algebra of functions
on $M$, of a Hilbert space $\H$ which is an analog of the space of square integrable spinors,
and of a unbounded operator $\DD$ which corresponds to a Dirac operator.
Since the algebra $\A$ is not required to be commutative,
one naturally incorporates noncommutative geometries in this approach.

Moreover, even the field theory actions in noncommutative geometry \cite{CC}
are defined through a trace of a function of $\DD$. Noncommutative geometry and
noncommutative field theory share common technique with quantum field theory.
This technique is the spectral geometry (see monographs \cite{Gilkey,Gilkey2,EliBook,KirBook,Vassilevich}).
This unity between different branches of science goes far beyond the pure technical level.
In a sense, noncommutative geometry is always quantum.

The algebra $\A$ need not be noncommutative. A lot of efforts were spent on analyzing
commutative and almost commutative spectral triples in the context of noncommutative
geometry, see \cite{ConnesMarcolli} for a overview. It has been realized that at the
scales much smaller than a certain cutoff scale $\Lambda$ the spectral action is
represented by an expansion, with all terms being integrals of local expressions
essentially coinciding with the heat kernel coefficients. Among that terms one can
find the Einstein--Hilbert action, the Yang--Mills action, the Higgs potential, and all
other ingredients of the Standard Model. Even the coefficients come out correctly,
so that one can speak now about checking phenomenological consequences of the
spectral action principle. We can say therefore that local aspects of (almost)
commutative spectral actions  are quite well understood. The study of global aspects,
including nonlocal terms, rapidly varying fields, etc. has been initiated
rather recently. These aspects is the main subject of the present paper.

We start below with a long pedagogical introduction in spectral triples
(Section \ref{sec-spectraltriple}) and spectral actions (Section \ref{Spectral action}).
We shall introduce the spectral dimension, the Wodzicki residue, the noncommutative integral,
and other useful notions. We describe in detail the large $\Lambda$ expansion
of the spectral action, which is essentially a generalization of the heat kernel expansion.
For the readers who do not want to go deep into mathematics we derive again this expansion
in the commutative case by rather elementary methods in Section \ref{commut}. This type of
the expansions assumes that the fields are small and slowly varying, i.e., they are weak-field
approximations. Few references on physical applications of spectral action are given in Section \ref{meaning} 
while the difference between the action and its asymptotic is recalled is Section \ref{aboutconvergence}. 
Section \ref{Duhamelexpansion} is devoted to the convergence of the Dyson--Phillips (also called Duhamel) 
expansion, namely, when a perturbation of a generator of the heat kernel gives rise, taking its trace, to 
an expansion series in terms of the perturbation, see \eqref{Trace}.

Various approximations to the spectral actions are based on corresponding expan\-sions of 
heat kernel. It is interesting therefore to construct an expansion of the heat kernel
in the fields (potential, curvatures, etc) without assuming that they vary slowly.
On the plain $\mathbb{R}^d$ this has been done long ago
\cite{Barvinsky:1987uw,Barvinsky:1990up} and recently translated into
expansions for the spectral action \cite{ILV1}. In Section \ref{pert} we consider
this problem on the torus $\mathbb{T}^d$, that is considerably more complicated due to
the presence or extra dimensionfull parameters -- circumferences of the torus.
We compute the heat kernel to the second order in the potential and gauge field
strength and consider various limits, including large and small proper time,
slowly and rapidly varying external fields. We also discuss implications for the
spectral action.

We dedicate this paper to Stuart Dowker on his 75th birthday.

\section{Notion of spectral triple}
\label{sec-spectraltriple}

The main properties of a compact spin Riemannian manifold $M$ can be recaptured using the following triple
$(\A=C^\infty(M), \H=L^2(M,S), \Ds)$. The coordinates $x=(x^1,\cdots,x^d)$ are exchanged with the algebra
$C^\infty(M)$, and the Dirac operator $\DD$ acting on the space $\H$ of square integrable
spinors
gives not only the dimension $d$ but also the metric of $M$ via
Connes formula and more generally generates a quantized calculus. The idea of noncommutative geometry is to
forget about the commutativity of the algebra and to impose axioms on a triplet $(\A,\H,\DD)$ to generalize the above
one in order to be able to obtain appropriate definitions of important notions: pseudodifferential operators, measure
and integration theory, etc.

This scheme is quite minimalist. Clearly, one needs the algebra of functions to be
able to talk about a manifold. The Dirac operator defines derivations and thus
length scales. The Hilbert space $\mathcal{H}$ is needed to give a domain for the
Dirac operator. Thus, the three ingredients of the triplet are necessary to define
a meaningful geometry and/or physics. As we shall see below, they are also sufficient.

The idea of the construction outlined below is to formulate basic properties
of the triple $(\mathcal{A},\mathcal{H},\slashed{D})$ and then use these properties
in a more abstract setting when commutativity of $\mathcal{A}$ is not assumed.
On this way one arrives at the following

\begin{definition}
\label{deftriplet}
A spectral triple $(\A,\H,\DD)$ is the data of an involutive (unital) algebra $\A$ with a faithful representation $\pi$ on
a Hilbert space $\H$ and a selfadjoint operator $\DD$ with compact resolvent
such that
$[\DD,\pi(a)]$ is bounded for any $a \in \A$.\end{definition}

In the usual commutative case compactness of the resolvent ensures that the
differential operator $\DD$ is no less than a first order operator. Boundedness
of the commutator means that $\DD$ is at most of the first order. This motivates
the definition above. This definition may be supplemented by more detailed
restrictions on the spectral triple.

The triple is even if there is a grading operator $\chi$ such that $\chi=\chi^*$,
$$
[\chi,\pi(a)]=0, \, \forall a \in \A\text{ and }\DD \chi=-\chi \DD.
$$
The operator $\chi$ mimics the usual chirality operator $\gamma_5$.

Properties of the gamma-matrices depend periodically on the dimensionality of space,
with the period equal to eight. Nonequivalent dimensions are distinguished by the
properties of the gamma's, and thus of the Dirac operator with respect to a
conjugation. These are precisely the properties which allow (or not) the existence
of Majorana, or Weyl, or Majorana--Weyl spinors. Being translated to the language
of noncommutative geometry this means that the spectral triple is
"real of KO-dimension $d \in \Z/8$" if there is an antilinear isometry $J:\H \to \H$ such that
$J\DD=\epsilon \,\DD J,\, J^2=\epsilon',\, J\chi=\epsilon'' \,\chi J$
with a table of signs $\epsilon, \epsilon',\epsilon''$ given in \cite{ConnesMarcolli,Polaris}
and the following commutation rules $[\pi(a),\pi(b)^\circ ]=0,\, \big[ [\DD,\pi(a)],\pi(b)^\circ \big]=0, \, \forall a,b\in \A
$
where $\pi(a)^\circ \vcentcolon = J \pi(a^*)J^{-1}$ is a representation of the opposite algebra 
\footnote{The opposite algebra $\A^\circ$ for $\A$ consists of the elements $a^\circ, a\in\A$ with a transposed 
multiplication, $a^\circ b^\circ =(ba)^\circ$.} 
$\A^\circ$.

The spectral triple is $d$-summable (or has metric dimension $d$) if the singular values $\mu_n$
\footnote{Singular values of $A$ are eigenvalues of $\vert A \vert$.}
of $\DD^{-1}$ behave for large $n$ as
$\mu_n(\DD^{-1})=\mathcal{O}(n^{-1/d})$. In the commutative case, this is just the
Weyl formula for distribution of eigenvalues.

They key result is the Reconstruction Theorem
\cite{CReconstruction,ConnesUnitary}, telling us that given a commutative spectral triple satisfying the above axioms 
and some more requirements, there exists a compact spin$^c$ manifold $M$ such that $\A \simeq C^\infty(M)$ 
and $\DD$ is a Dirac operator.

Let us continue with constructing relevant objects of noncommutative geometry through
the spectral triple. The operator $\DD$ tells us what a derivative is and thus generates
naturally the set of pseudodifferential operators $\Psi(\A)$ \cite{ConnesMarcolli}.
For $P \in \Psi(\A)$, we define the zeta-function associated to $P$ (and $\DD$) by
\begin{align}
\label{zetaPD}
\zeta_\DD^P: \, s \in \C \to \Tr \big(P \DDD^{-s} \big)
\end{align}
which makes sense since for $\Re(s) \gg 1$, $P \DDD^{-s}$ is trace-class.

The dimension spectrum $Sd$ of the triple is the set of poles of
$\zeta_\DD^P(s) \, \forall P \in \Psi(\A)$. It is said simple if the poles have order at most one.

\begin{prop}
\label{spectrcomm}
Let $Sd(M)$ be the dimension spectrum of a commutative geometry of dimension $d$. Then $Sp(M)$ is simple and
$Sd(M) = \set{d-k \, \vert  \, k \in \N}$.
\end{prop}

Finally, one can define a trace, called a noncommutative integral of $P$, that is given by
\begin{align*}
\ncint P \vc \underset{s=0}{\Res} \,\zeta_\DD^P(s).
\end{align*}
In \eqref{zetaPD}, we assume $\DD$ invertible since otherwise, one can replace $\DD$ by the invertible operator
$\DD + {\rm Pr}$, ${\rm Pr}$ being the projection on $\Ker\DD$.
As noticed by Wodzicki, $\ncint P$ is equal to $-2$ times the coefficient in log $t$ of the asymptotics of
$\Tr(P\, e^{-t\,\Ds^2}$) as $t \rightarrow 0$. It is remarkable that this coefficient is independent of $\Ds$
and this gives a close relation between the $\zeta$ function and
heat kernel expansion with the Wodzicki residue {\it WRes} \cite{Wodzicki,Wodzicki1}. Actually,
\begin{align*}
\Tr(P\, e^{-t\,\Ds^2}) \underset{t \downarrow 0^+}{\sim} \, \sum_{k=0}^{\infty} a_k\,t^{(k-ord(P)-d)/2}
+ \sum_{k=0}^{\infty} (-a'_k\,\log t+b_k)\,t^k,
\end{align*}
so $\ncint P=2 a'_0$. 

Since $\ncint$ and {\it WRes} are traces on $\Psi\big(C^\infty(M)\big)$, thus by uniqueness 
$\ncint P=c \,\text{\it WRes }P$. The Wodzicki residue is known in quantum field theory due to
its' relation to the multiplicative anomaly \cite{Elizalde:1997nd}.

Let us proceed with differential forms.
The algebra $\mathcal{A}$ gives us smooth functions, and thus the $0$-forms.
The set of one-forms may be defined as \cite{Book,Polaris}
\begin{align*}
\Omega^1_\DD(\A) \vc \text{span}\set{ a \, db\, \,\vert\, \, a,b \in \A}, \quad db \vc [\DD,b].
\end{align*}
In the commutative case, this defines 1-forms contracted with $\gamma$-matrices
rather than usual forms.
One can add a one-form $A$ to $\DD$ to get $\DD_A:=\DD+A$, but when a reality operator $J$ exists, we also want
$\DD_A J=\epsilon \, J \DD_A$, so we choose
\begin{align}
\label{fluct}
\DD_{\wt A} \vc \DD + \wt A, \quad \wt A \vc A + \epsilon JAJ^{-1}, \quad A=A^*.
\end{align}
In the commutative case, $\A^\circ \simeq J\A J^{-1} \simeq \A$ which also gives
\begin{align*}
JAJ^{-1}=-\epsilon\,A^* , \quad \forall A\in \Omega^1_\DD(\A) \text{ thus } \wt A=0 \text{ when } A=A^*.
\end{align*}
This does not mean that $\DD$ cannot fluctuate in commutative geometries, but that one has to consider only 
non-symmetrized fluctuations $\DD_A$.

\section{Spectral action}
\label{Spectral action}

We would like to obtain a good action for any spectral triple and for this it is useful to look at some examples in
physics. In any physical theory based on geometry, the interest of an action functional is, by a minimization
process, to exhibit a particular geometry, for instance, trying to distinguish between different metrics.
This is the case in general relativity with the Einstein--Hilbert action (with its Riemannian signature).

The Einstein--Hilbert action is
\begin{align*}
S_{EH}(g)\vc-\int_M R_g(x) \, dvol_g(x)
\end{align*}
where $R$ is the scalar curvature (chosen positive for the
sphere). This is nothing else in dimension 4 (up to a constant) than $\ncint \Ds^{-2}$.

But in the search for invariants under diffeomorphisms, there are more quantities than the
Einstein--Hilbert action, a trivial example being $\int_M f\big(R_g(x)\big) \,dvol_g(x)$, and there are others.
In this desire to implement gravity in noncommutative geometry, the eigenvalues of the Dirac
operator look as natural variables \cite{LR}.
However we are looking for invariants which add up under disjoint unions of different geometries.

\subsection{Quantum approach and spectral action}
In a way, a spectral triple fits quantum field theory since $\DD^{-1}$ can be seen as the propagator for (Euclidean) 
fermions and we can compute Feynman graphs with fermionic internal lines. The gauge
bosons are only derived objects obtained from internal fluctuations
described by a choice of a connection which is associated to a one-form in $\Omega_\DD^1(\A)$.
Thus, the guiding principle followed by Chamseddine and Connes
is to use a theory which is of pure geometric origin
with a functional action based on the spectral triple, namely which depends only on the spectrum of $\DD$ \cite{CC}.
They proposed the following

\begin{definition}
The spectral action of a spectral triple $(\A,\H,\DD)$ is defined by
\begin{align}
\label{defsa}
\SS(\DD,f,\Lambda)&:=\Tr \big(f(\DD^2/\Lambda^2) \big) 
\end{align}
where $\Lambda\in \R^+$ plays the role of a cut-off and $f$ is any positive function (such that $f(\DD^2/\Lambda^2)$
is a trace-class operator).
\end{definition}
One can also define as in \cite{ConnesMarcolli} $\SS(\DD,g,\Lambda)=\Tr \big(g(\DD/\Lambda) \big)$ where 
$g$ is positive and even. With this second definition, $S(\DD,g,\Lambda) = \Tr \big(f(\DD^2/\Lambda^2)\big)$ 
with $g(x)\vc f(x^2)$.

As an example
for $f$, one can take the characteristic function of $[0,1]$. Then
$\Tr \big( f(\DD^2/\Lambda^2)\big)$ is
nothing else but the number of eigenvalues of $\DD$ within $[-\Lambda,\Lambda]$.

When this action has an asymptotic series in $\Lambda \to \infty$, we deal with an effective theory.
Naturally, $\DD$ has to be replaced by $\DD_A$ which is a just a decoration. To this bosonic part of the action,
one adds a fermionic term $\tfrac{1}{2}\langle J \psi,\DD \psi\rangle$ for $\psi\in \H$ to get a full action.
In the standard model of particle physics, this latter corresponds  to the
integration of the Lagrangian part for the coupling of gauge bosons and Higgs bosons with fermions.

\subsubsection{Yang--Mills action}
\label{YMA}

This action plays an important role in physics so it is natural to consider it in the
noncommutative framework. Recall first the classical situation: let G be a compact Lie group with its Lie algebra
$\mathfrak{g}$ and let $A\in \Omega^1(M,\mathfrak{g})$ be a connection. If
$F \vc da+ \tfrac{1}{2}[A,A] \in \Omega^2(M,\mathfrak{g})$ is the curvature (or field strength) of $A$, then the
Yang--Mills action is $S_{YM}(A)=\int_M \tr(F \wedge \star F)$.  In the abelian
case $G=U(1)$, it is the Maxwell action and its quantum version is the quantum electrodynamics (QED) since the
un-gauged $U(1)$ of electric charge conservation can be gauged, and its gauging produces electromagnetism
\cite{Schucker}. It is conformally invariant when the dimension of $M$ is $d=4$.

The Yang--Mills action can be defined in the context of noncommutative geometry
for a spectral triple $(\A,\H,\DD)$ of dimension
$d$ \cite{Connes88,Book}. Let $A \in \Omega_\DD^1(\A)$ and curvature $\theta=dA+A^2$.
Then it is natural to consider
$$
I(A)\vc \Tr_{Dix}(\theta^2 \DDD^{-d})
$$
since it coincides (up to a constant) with the previous Yang-Mills action in the commutative case. Here $\Tr_{Dix}$
is the Dixmier (singular) trace \cite{Dixmier}:  if $P=\theta^2 \vert \DD \vert^{-d}$, for the principal symbol,
$\tr\big(\sigma^P(x,\xi)\big)=c\,\tr(F \wedge \star F)(x)$, and the Dixmier trace is also related to Wodzicki
residue for PDO's on compact manifolds (uniqueness of traces up to a constant).
The key observation regarding the Yang--Mills action is that it appears in the
$1/\Lambda$ expansion of the spectral action, see Sec.\ \ref{commut}.

The spectral action is more conceptual than the Yang--Mills action since it gives no fundamental role to the
distinction between gravity and matter in the artificial decomposition $\DD_A=\DD+A$. For instance, for the minimally
coupled standard model, the Yang--Mills action for the vector potential is part of the spectral action, as well as the
Einstein--Hilbert action for the Riemannian metric \cite{CC2}.

As quoted in \cite{CC4}, the spectral action has conceptual advantages:

- Simplicity: when $f$ is a cutoff function, the spectral action is just the counting function.

- Positivity: when $f$ is positive (which is the case for a cutoff function), the action
$\Tr \big( f(\DD^2/\Lambda^2)\big) \geq 0$ has the correct sign for a Euclidean action: the positivity of the function
$f$ will insure that the actions for gravity, Yang-Mills, Higgs couplings are all positive and the Higgs mass term
is negative.

- Invariance:  the spectral action has a much larger invariance group than the usual diffeomorphism group as for
the gravitational action; this is the unitary group of the Hilbert space $\H$.

However, this action is not local. It only becomes so when it is replaced by the asymptotic expansion:

\subsection{Asymptotic expansion for $\Lambda \to \infty$}
\label{asymptoticspectralaction}

The heat kernel method already used in previous sections will give a control of spectral action
$S(\DD,f,\Lambda)$ when $\Lambda$ goes to infinity.

\begin{theorem} \cite{ConnesMarcolli}
Let $(\A,\H,\DD)$ be a spectral triple with a simple dimension spectrum $Sd$.
\\
We assume that
\begin{align}
\label{hyp}
\Tr\big(e^{-t\DD^2}\big)  \, \underset{t \,\downarrow \,0}{\simeq}  \,\sum_{\a\in Sd} a_\a\,t^\a.
\end{align}

(i) If $a_\a\neq 0$ with $\a<0$, then the zeta function $\zeta_\DD$ defined in \eqref{zetaPD} has a pole at $-2\a$ with
$\underset{s=-2\a}{\Res} \, \zeta_\DD(s)=\tfrac{2a_\a}{\Gamma(-\a)}$.

(ii) For $\a=0$, we get $\zeta_\DD(0) =a_0-\dim \,\Ker \,\DD$.

(iii)One has the following  asymptotic expansion over the positive part $Sd^+$ of $Sd$:
\begin{align}
\label{asympspectral}
\Tr\big(f(\DD/\Lambda)\big)  \, \underset{\Lambda \,\to+\infty}{\simeq}  \,
\sum_{\beta \in Sd^+} f_\beta\,\Lambda^\beta \ncint \vert \DD \vert^{-\beta}+ f(0)\,\zeta_\DD(0)+ \cdots
\end{align}
where the dependence of the even function $f$ is $f_\beta:=\int_0^\infty f(x)\,x^{\beta-1}\,dx$ and $\cdots$ involves
the full Taylor expansion of $f\,$at 0.
\end{theorem}
Here, one assumes sufficient hypothesis on $f$ like $f$ is a Laplace transform with $\vert f_\beta \vert<\infty$, etc.
It can be useful to make a connection between the spectral action and heat expansion \cite{Gilkey,Gilkey2}:

\begin{corollary}
Assume that the spectral triple $(\A,\H,\DD)$ has dimension $d$.
\label{asymptidentic}

If ${\rm Tr} \big( e^{-t \, \DD^2} \big) \, \underset{t \,\downarrow \,0}{\simeq}  \,
\sum_{k\in \set{0,\cdots,d}} t^{(k-d)/2} \,{a}_k (\DD^{2})+\cdots$, then
\begin{align}
\label{asymp1}
\SS(\DD,f,\Lambda)\underset{t \,\downarrow \,0}{\simeq}  \,
 \sum_{k \in \set{1,\cdots ,d}} \,f_k \, \Lambda^k\,a_{d-k}(\DD^2) +f(0)\,a_d(\DD^2)+\cdots
\end{align}
with $f_k:=\tfrac{1}{\Ga(k/2)}\int_0^\infty f(s)s^{k/2-1}ds$. Moreover,
\begin{align}
& a_k(\DD^2)=\tfrac 12\,\Gamma(\tfrac{d-k}{2})\,\ncint \vert\DD \vert^{-d+k} \text{ for }k=0,\cdots,d-1, \label{nlccoeff}.\\
& a_d(\DD^2)= \dim\, \Ker \DD+\zeta_{\DD^2}(0). \nonumber
\end{align}
\end{corollary}

The spectral action uses the value of $\zeta_\DD(0)$ in the constant term $\Lambda^0$ of the asymptotics
\eqref{asympspectral}. So it is fundamental to look at its variation under a gauge fluctuation $\DD \to \DD+A$.
For instance $\zeta_{D_{\wt A}}(0)=\zeta_{D}(0)+\sum_{q=1}^{d} \tfrac{(-1)^{q}}{q} \ncint (\wt AD^{-1})^{q}$, 
\cite{CC1,EILS}

\subsection{Remark on the use of Laplace transform}
\label{useofLaplace}

In \eqref{asymp1}, the spectral action asymptotic behavior
\begin{align}
\label{asympto}
S(\DD,f, \Lambda) \, \underset{\Lambda \,\to+\infty}{\simeq}  \,  \sum_{n=0}^\infty c_n \,\Lambda^{d-n}\,a_n(\DD^2)
\end{align}
has been proved for a smooth function $f$ which is a Laplace transform for an arbitrary
spectral triple (with simple dimension spectrum) satisfying \eqref{hyp}. 
However, this hypothesis is too restrictive since it does not cover the heat kernel case where $f(x)=e^{-x}$.

When the triple is commutative and $\DD^2$ is a generalized Laplacian on sections of a vector bundle over a
manifold of dimension $d$, hypothesis \eqref{hyp} can be proved \cite{Gilkey} since the spectrum dimension is given by 
Proposition \ref{spectrcomm} and previous asymptotics \cite{EGBV} for $d=4$ is (see also next section)
\begin{align*}
\Tr \big( f(\DD^2/\Lambda^2) \big) &\simeq \tfrac{1}{(4\pi)^2} \Big[ \big(\text{rk(E)}\int_0^\infty  xf(x)\, dx \big)
\, \Lambda^4 + \big(b_2(\DD^2) \int_0^\infty f(x)\,dx\big)\, \Lambda^2  \\
&\hspace{3cm} +\sum_{m=0}^\infty \big((-1)^m\, f^{(m)}(0)\,b_{2m+4}
(\DD^2) \big) \, \Lambda^{-2m} \Big],\quad \Lambda \to \infty
\end{align*}
where $(-1)^m b_{2m+4}(\DD^2) = \tfrac{(4\pi)^2}{m!}  \,\mu_m(\DD^2)$ are suitably normalized, integrated
moment terms of the spectral density of $\DD^2$.

The main point is that {\it this asymptotics makes sense in the Ces\`aro sense }(see \cite{EGBV} for definition) for
$f$ in $\mathcal{K'}(\R)$, which is the dual of $\mathcal{K}(\R)$. This latter is the space of smooth functions
$\phi$ such that for some $a\in\R$, $\phi^{(k)}(x)=\mathcal{O}( \vert x\vert^{a-k} )$ as $\vert x \vert \to \infty$,
for each $k\in \N$. In particular, the Schwartz functions are in $\mathcal{K}(\R)$ (and even dense).

Of course, the counting function is not smooth but is in $\mathcal{K'}(\R)$, so such behavior \eqref{asympto}
is wrong beyond the first term, but is correct in the Ces\`aro sense. Actually there are more terms with
$\delta$'s and derivatives of $\delta$ as explained on examples in \cite[p. 243]{EGBV}.

\subsection{Commutative case}\label{commut}
It is instructive to re-derive asymptotics of the spectral action in a commutative
case where one does not need any of the new notions, like the Wodzicki residue or
the dimension spectrum. Let us take a compact spin manifold $M$ of dimension $d=4$
without a boundary and
$\DD=\Ds=i\ga^\mu(\nabla_\mu+A_\mu)$, being the standard Dirac operator with a
spin-connection $\nabla$ and a gauge field $A$. Let us suppose that $f$ is a Laplace
transform,
\begin{equation*}
f(z)=\int_0^\infty dt\, e^{-tz} \varphi (t)\,.
\end{equation*}
We also have an asymptotic expansion
\begin{equation}
\Tr \big( e^{-t\DD^2} \big) \, \underset{t\downarrow 0}{\sim} \,\sum_{k=0}^\infty t^{-2+k}\,a_{2k}(\DD^2).
\label{asymp}
\end{equation}
Therefore, neglecting all subtleties arising when commuting sums, traces and integrals,
we can write
\begin{eqnarray}
&&\SS (\DD,f,\Lambda)=\int_0^\infty dt \Tr \big( e^{-t\DD^2} \big) \, \varphi (t)
\simeq \int_0^\infty dt \sum_{k=0}^\infty t^{-2+k}\,a_{2k}(\DD^2) \varphi (t)\nonumber\\
&&\qquad\qquad \simeq \sum_{k=1}^\infty \Lambda^{2(2-k)} \,
\varphi_{2k} \, a_{2k}(\DD^2)\,,\label{commutas}
\end{eqnarray}
where
\begin{equation}
\varphi_{2k}=\int_0^\infty dt\, t^{-2+k}\varphi (t) \,.\label{phi2k}
\end{equation}
The heat kernel coefficients for $\DD^2$ may be found in a textbook, see \cite[Sec.4.4]{FV}:
\begin{eqnarray*}
&&a_0(\DD^2)= \frac 1{4\pi^2}\int_M d^4x \sqrt{g}\, {\rm tr}\, (1),\\
&&a_2(\DD^2)= -\frac 1{48\pi^2}\int_M d^4x \sqrt{g}\, {\rm tr}\, (R),\\
&&a_4(\DD^2)= \frac 1{24\pi^2}\int_M d^4x \sqrt{g}\, {\rm tr}\, \big(-F_{\mu\nu}F^{\mu\nu}
+(R^2\mbox{-terms})\big),
\end{eqnarray*}
where the trace is taken over gauge indices and
$F_{\mu \nu}=\partial_\mu A _\nu-\partial_\nu A_\mu+ [A_\mu,A_\nu]$. Thus we obtained
an expansion of the spectral action in $1/\Lambda^2$, where we can find many relevant terms.
The $a_0$ term is the cosmological term, while $a_2$ is the Einstein--Hilbert action.
The $a_4$ term contains the Yang--Mills action and curvature-squared corrections to the
gravity action. Note that the Einstein--Hilbert and Yang--Mills actions came out with
correct signs (assuming $\varphi_{2k}$ positive).

Each of the heat kernel coefficients $a_{2k}$ is an integral of a polynomial of the
canonical dimension $2k$, i.e., it contains a limited number of fields and derivatives.
Therefore, we are dealing with a \emph{weak-field expansion} of the spectral action.

In \cite{vanSuijlekom:2011kc} it was suggested to choose a function $f$ such that the
expansion \eqref{asympto} or \eqref{ncomS}  contains a finite number of terms, and to use this expansion
instead of the full spectral action. Although such a theory may have some nice properties,
they do not carry over to the full theory \cite{ILV1}.

\subsection{About the meaning of the spectral action via its asymptotics}
\label{meaning}

We have discussed above the spectral action for compact spectral triples. In a non-compact
case, which is more physically motivated frequently, the definition of spectral action
has to be modified \cite{ILV1}
\begin{equation}
\SS(\DD,f,\Lambda):=\Tr \big(f(\DD^2/\Lambda^2) - f(\DD_0^2/\Lambda^2)\big)\,,\label{ncomS}
\end{equation}
where $\DD_0$ is the unperturbed Dirac operator with $A=0$.

Almost commutative geometry, which is a commutative geometry times a finite one (where the finite one is a sum 
of matrices) has been deeply and intensively investigated for the noncommutative
approach to the standard model of particle physics, see \cite{CCM,ConnesMarcolli}. This approach offers a
lot of interesting perspectives. For instance, the possibility to compute the Higgs representations and mass
(for each noncommutative model) is
particularly instructive \cite{CC,CC6,CC9,LMM,JS,JS1,MGV}. The choice of the Dirac
operator is dictated by coupling of the fermions in standard model. It is interesting,
that the whole bosonic action is then reproduced automatically.

The spectral action has been computed in \cite{ILS} for the quantum group $SU_q(2)$ which is not a
deformation of $SU(2)$ of the type considered on the Moyal plane (\cite{GGBISV,GI,GIVas}).
It is quite peculiar since \eqref{asympspectral} has only a finite number of terms. Also on the Moyal
spaces, the heat kernel expansion and the expansion (\ref{asympto}) of the spectral action
was calculated in a number of papers \cite{GGBISV,GI,GIVas,Vm1,Vm2,Vm3}. The properties of
both expansions depend crucially on the number of compact noncommutative dimensions,
as discussed in \cite{SV}.

Due to the difficulties to deal with non-compact manifolds,
the case of spheres $\mathbb{S}^4$ or $\mathbb{S}^3\times \mathbb{S}^1$ has been investigated in
\cite{CC4,CC5} for instance in the case of  Robertson-Walker metrics.

An approach to the spectral action based on quantum anomalies has been suggested in
\cite{Andrianov:2010nr} and applied to Higgs-dilaton interactions in \cite{Andrianov:2011bc}.

All the machinery of spectral geometry has been recently applied to cosmology, computing the spectral action in
few cosmological models related to inflation, see \cite{KM, MPT1,MPT2, MP,NS,Sa}.

Spectral triples associated to manifolds with boundary have been considered in \cite{CC7,CC8,IL1,IL2,ILV}. The
main difficulty is precisely to put nice boundary conditions to the operator $\DD$ to still get a selfadjoint
operator and then, to define a compatible algebra $\A$. This is probably a must to obtain a result in a
noncommutative Hamiltonian theory in dimension 1+3.

The case of manifolds with torsion has also been studied in \cite{HPS,PS,PS1}, and even with boundary
in \cite{ILV}. These works show that the Holst action appears in spectral actions and that torsion could
be detected in a noncommutative world.

Somewhat similar ideas that the gravity is a low-energy effect of quantized matter field
rather than a fundamental force was suggested long ago by Zeldovich \cite{Zel} and
Sakharov \cite{Sak}, see \cite{NV} for a review. The spectral action approach extends much
further: the spectral action is valid for all energies. This is why one is interested
in the effects which are not seen at the asymptotic expansions.

\subsection{About convergence and divergence, local and global aspects of the asymptotic expansion}
\label{aboutconvergence}

The asymptotic expansion series \eqref{asymp1} of the spectral action may or may not converge. It is known 
that each function $g(\Lambda^{-1})$ defines at most a unique expansion series when $\Lambda \to \infty$ 
but the converse is not true since several functions have the same asymptotic series. 
We give here examples of convergent and divergent series of this kind.

When $M$ is the torus $\T^d$ as in Section \ref{generalcase} with $\Delta=g^{\mu\nu} \partial_\mu\partial_\nu$, 
\begin{align*}
\Tr(e^{t\Delta})=\frac{(4\pi)^{-d/2} \,\Vol(T^d)}{t^{d/2}} +\mathcal{O}(t^{-d/2}\,e^{-1/4t}),
\end{align*}
thus the asymptotic series $\Tr(e^{t\Delta}) \simeq \frac{(4\pi)^{-d/2} \,\Vol(T^d)}{t^{d/2}}$, $t\to 0$, has only one term. 

In the opposite direction, let now $M$ be the unit four-sphere $\mathbb{S}^4$ and $\Dslash$ be the usual Dirac 
operator. By Propostion \ref{spectrcomm}, equation \eqref{hyp} yields (see \cite{CC5}):
\begin{align*}
& \Tr(e^{-t\Dslash^2})=  \frac{1}{t^2} \big(  \frac{2}{3} + \frac{2}{3} \,t 
+ \sum_{k=0}^n a_k \,t^{k+2} +\mathcal{O}(t^{n+3}) \big) , \\
& a_k:=\frac{(-1)^k \,4}{3\, k!} \big( \frac{B_{2k+2}}{2k+2}-\frac{B_{2k+4}}{2k+4} \big)
\end{align*}
with Bernoulli numbers $B_{2k}$. 
Thus $ t^2\Tr(e^{-t\Dslash^2})\simeq \frac{2}{3} + \frac{2}{3} \,t + \sum_{k=0}^\infty a_k \,t^{k+2}$ when 
$t \to 0$ and this series is a not convergent but only asymptotic: 
$a_k > \tfrac{4}{3\,k!}\tfrac{\vert B_{2k+4}\vert }{2k+4}  > 0$ and $\vert B_{2k+4} \vert 
=2\, \frac{(2k+4)!}{(2\pi)^{2k+4}} \, \zeta(2k+4) \simeq 4 \sqrt{\pi (k+2)} 
\left(\frac{k+2}{ \pi e} \right)^{2k+4}  \to \infty \text{ when }k \to \infty$.

More generally, in the commutative case considered above and when $\DD$ is a differential operator---like a 
Dirac operator, the coefficients of the asymptotic series of $\Tr(e^{-t\DD^2})$ are locally defined by the symbol of 
$\DD^2$ at point $x\in M$ but this is not true in general: in \cite{GilkeyGrubb} is given a positive elliptic 
pseudodifferential such that non-locally computable coefficients especially appear in \eqref{asymp} when $2k> d$. 
Nevertheless, all coefficients are local for $2k \leq d$.

Recall that a locally computable quantity is the integral on the manifold of a local frame-independent smooth 
function of one variable, depending only on a finite number of derivatives of a finite number of terms in the 
asymptotic expansion of the total symbol of $\DD^2$. 
For instance, some nonlocal information contained in the ultraviolet asymptotics can be recovered if one looks at 
the (integral) kernel of $e^{-t\sqrt{-\Delta}}$: in $\T^1$, with $\Vol(\T^1)=2\pi$, we get \cite{Fulling}
$$
\Tr(e^{-t\sqrt{-\Delta}})=\frac{\sinh (t)}{\cosh(t)-1}=\coth(\frac{t}{2})=
\frac{2}{t}\,\sum_{k=0}^\infty \frac{B_{2k}}{(2k)!} \,t^{2k}=\frac{2}{t}[1+\frac{t^2}{12}-\frac{t^4}{720} +\mathcal{O}(t^6)]
$$
and the series converges when $t<2\pi$, since $\frac{B_{2k}}{(2k)!}=(-1)^{k+1} \,\tfrac{2\,\zeta(2k)}{(2\pi)^{2k}}$, thus 
$\tfrac{\vert B_{2k} \vert}{(2k)!} \simeq \tfrac{2}{(2\pi)^{2k}}$ when $k\to \infty$.

Thus we have an example where $t \to \infty$ cannot be used with the asymptotic series.

Thus the spectral action \eqref{asymp1} precisely encodes these local and nonlocal behavior which appear or not
in its asymptotics for different $f$. The coefficient of the action for the positive part (at least) of the dimension 
spectrum correspond to renormalized traces, namely the noncommutative integrals of \eqref{nlccoeff}. 
In conclusion, the asymptotic \eqref{asympto} of spectral action may or may not have nonlocal coefficients.

For the flat torus $\T^d$, the difference between $\Tr(e^{t\Delta})$ and its asymptotic series is an oscillatory term 
which is related to periodic orbits of the geodesic flow on $\T^d$. Similarly, the counting function $N(\lambda)$ 
(number of eigenvalues including multiplicities of $\Delta$ less than $\lambda$) obeys Weyl's law: 
$N(\lambda)= \frac{(4\pi)^{-d/2} \,\Vol(\T^d)}{\Gamma(d/2 +1)}\, \lambda^{d/2} + o(\lambda^{d/2})$ --- see \cite{Arendt} 
for a nice historical review on these fundamental points.
The relationship between the asymptotic expansion of the heat kernel and the formal expansion of the spectral 
measure is clear: the small-$t$ asymptotics of heat kernel is determined by the large-$\lambda$ asymptotics of the 
density of eigenvalues (and eigenvectors). However, the latter is defined modulo some average: 
Ces\`aro sense as reminded in Section \ref{useofLaplace}, or Riesz mean of the measure which washes out 
ultraviolet oscillations, but also gives informations on intermediate values of $\lambda$ \cite{Fulling}.

In \cite{CC4,MPT1} are given examples of spectral actions on (compact) commutative geometries of dimension 4 
whose asymptotics have only two terms. In the quantum group $SU_q(2)$, the spectral action \eqref{defsa} 
itself has only 4 terms, independently of the choice of function $f$.

\section{Trace-class convergence of Dyson--Phillips series (Duhamel expansion)}
\label{Duhamelexpansion}

We review here a few facts about the Gibbs semigroups, the Duhamel or Dyson--Phillips expansion and
related convergence questions. We refer to \cite{Kato,BA,ZagrebnovBook} for more information on this subject.

A \emph{$C_0$-semigroup} (or \emph{strongly continuous semigroup}) is a family $(G(t))_{t\in \R_+}$
of bounded operators on a Hilbert space $\H$, such that, $G(0)=\Id$, for any $t,t'\in \R_+ $, $G(t+t')=G(t)G(t')$, and
$t\mapsto G(t)$ is a continuous map in the strong operator topology sense, or in other words,
$(G(t))(\psi)$ is a continuous function of $t$ for any fixed $\psi\in \H$.
\\
Given a $C_0$-semigroup $(G(t))_{t\geq 0}$, the \emph{generator} of the semigroup is the operator
$T$ on $\H$, defined on $\Dom T:=\set{\psi \in \H \ : \ T\psi:=\lim_{h\to 0^+} h^{-1}\big(G(h)\psi-\psi\big)\ \text{exists}}$. 
It turns out that $T$ determines the semigroup uniquely and is a closed densely defined operator on $\H$.
Moreover, if $T$ is selfadjoint, $G(t)=e^{tT}$, where $e^{tT}$ is defined thanks to the spectral theorem.
When $T$ generates $(G(t))_{t\geq 0}$, the map $u:t\to G(t)\,\psi$ on $\R_+$ for a given $\psi\in \H$, is the
unique solution to the following abstract Cauchy problem:
\begin{align*}
u'(t)= T \,u(t)\,, \quad u(0)=\psi \, .
\end{align*}
It is therefore not surprising that operator semigroup theory is particularly useful for the description
of phenomena associated to linear evolution equations.

A good part of this theory is devoted to the \emph{generation problem}, which is
finding conditions on a closed densely defined operator $T$ so that it is the generator
of a semigroup with given desired properties. The general case is solved by the Feller--Miyadera--Phillips theorem:
Fix $w\in \R,\,M\in \R^+$ and let $T$ be a closed densely defined operator such that for for any
$\la\in \C_{>w}:=\set{\mu \in \C \ :\ \Re(\mu)>w}$,
the number $\la$ is in the resolvent set of $T$ and the following estimate holds for any $n\in \N$:
$$
\norm{(T-\la)^{-n}}\leq M(\Re(\la)-w)^{-n}\, .
$$
Then, $T$ is the generator of a $C_0$-semigroup $(G(t))_{t\geq 0}$ satisfying $\norm{G(t)}\leq Me^{wt}$ for all
$t\geq 0$.
Since \emph{any} $C_0$-semigroup satisfies an estimate of this type, this result provides a general description
of all possible semigroup generators.
This theorem is a generalization of the Hille--Yosida theorem, which concerns the generation of
\emph{contractive} semigroups ($\norm{G(t)}\leq 1,\,\forall t\in \R_+$), and is obtained by considering $w=0$ and
$M=1$ in the previous formulation.

As an application, one can show that any selfadjoint operator $T$ generates a $C_0$-semigroup if and only if it
is bounded above. For example, any selfadjoint bounded perturbation of minus the Laplacian on $\R^n$,
acting on $L^2(\R^n)$, yields a $C_0$-semigroup.

If the operator $H=-T$ is interpreted as an unperturbed Hamiltonian, the operator $H+P$ obtained by
a suitable perturbation $P$ can be seen as an Hamiltonian of an interacting system.
This fact motivates the \emph{perturbation problem} of operator semigroup theory: given a generator $T$ of a
strongly continuous semigroup, what are the conditions on an operator $B$ so that $T+B$ is the generator of a
strongly continuous semigroup ?

It turns out that any bounded perturbation $B$ of a generator $T$ of a $C_0$-semigroup $(G_T(t))_{t\geq 0}$
is a generator of a strongly continuous semigroup $(G_{T+B}(t))_{t\geq 0}$. Moreover, the perturbed semigroup
$G_{T+B}$ can be obtained as a  \emph{Dyson--Phillips series} (also called \emph{Duhamel expansion}):
\begin{align}
\label{DDP}
G_{T+B}(t) = \sum_{n=0}^\infty G_n(t), \quad t\geq 0
\end{align}
where the sequence of operators $(G_n)$ is inductively defined by $G_0(t)=G_A(t)$ and
$$
G_{n+1}(t):= \int_0^t G_{T}(t-s) \,B\,G_n(s)\,ds
$$
in the strong operator sense. The convergence of \eqref{DDP} is here to be understood in the
norm topology (uniform convergence), and is obtained by iterative application of the \emph{Duhamel formula}:
\begin{align*}
G_{T+B}(t)-G_T(t)= \int_0^t G_{T+B}(t-s) \,B \,G_T(s)\, ds\, , \quad t \geq 0
\end{align*}
in the strong operator sense.

Many natural semigroups $(G(t))$ that appear in quantum statistical mechanics or in heat kernel theory
are actually families of operators which are, not only bounded, but also trace-class when the parameter $t$ is nonzero.
This means that for any $t>0$, $\norm{G_t}_1:=\sum_{k\in \N}|\langle e_k, G_t e_k\rangle|$ is finite and the  trace
of $G_t$  exists:  $\Tr(G_t)= \sum_{k\in \N} \langle e_k, G_t e_k\rangle$, where $(e_k)$ is any orthonormal basis of
$\H$.
A strongly continuous semigroup which has this property is called a \emph{Gibbs semigroup} \cite{Uhlenbrock}.
It turns out that the condition of finiteness of the trace-norm $\norm{G(t)}_1$ for each $t>0$ automatically implies
continuity of the map $t\to G(t)$ in the topology of the $\norm{\cdot}_1$-norm \cite[Proposition 2]{Uhlenbrock}.

The natural question related to perturbation theory is now in this setting: can we extend the Dyson-Phillips
expansion formula \eqref{DDP} to the trace of $G_{T+B}(t)$ and $G_n(t)$?
This question has been answered positively by Uhlenbrock \cite[Theorem 3.2]{Uhlenbrock}.
He proved that if $G_T$ is a Gibbs semigroup with generator $T$ and if the perturbation $B$ is bounded,
then the perturbed $C_0$-semigroup $G_{T+B}$ is a Gibbs semigroup, and the Dyson--Phillips series \eqref{DDP}
converges in the trace-norm $\norm{\cdot}_1$ sense. This implies in particular that for all $t>0$,
\begin{align}
\label{Trace}
\Tr (G_{T+B}(t)) =
\sum_{n=0}^\infty \int_{R^t_n} \Tr\big(G_T(t-s_{n-1})B G_{T}(s_{n-1}-s_{n-2})
\cdots G_T(s_1-s_0) B G_T(s_0)\big)ds
\end{align}
where $R_n^t:=\set{(s_0,\cdots,s_{n-1}) \ : \ 0\leq s_0\leq \cdots\leq s_{n-1}\leq t}$.

In some physical situations, the boundedness condition on the perturbation $B$ is too strong. The question
on possible generalization to unbounded perturbation has been first investigated in \cite{ANB}, where the trace-norm
convergence of the Dyson--Phillips series has been obtained for perturbations in the $\mathcal{P}_0$-class,
that is for closed operators $B$ with domain containing
$\bigcup_t \,G_T(t) (\H)$, and such that $\int_0^1 \norm{BG_T(t)}\, dt <\infty$.
This class is actually included in the set of operators which are \emph{relatively bounded with respect to}
$T$ with relative bound equal to 0 \cite[Theorem 2.2]{Zagrebnov1}.
Recall that an operator $B$ is relatively bounded to $T$ (or $T$-bounded) if the domain of $B$
contains the domain of $T$ and $\norm{B\psi}\leq a \norm{\psi} + b \norm{T\psi}$ for all $\psi$ in the domain
of $T$, for constants $a,b\geq 0$. The infimum of all possible values of $b$ in previous estimate is called
the \emph{relative bound} of $B$ with respect to $T$ \cite[p. 190]{Kato}.

As classical examples of perturbation operators $B$ with zero relative bound with respect to the Laplace operator
$\Delta$ on $\R^n$, one can consider the Kato--Rellich class of potential $V\in L^p(\R^n)+L^\infty(\R^n)$,
where $p=2$ if $n=3$ and $p> n/2$ if $n\geq 4$.
Note also that any first order differential operator is $\Delta$-bounded with zero relative bound.

The trace-norm convergence (as well as other analyticity questions) of the Dyson--Phillips expansion in the case of
a nonzero relative bound has been investigated by Zagrebnov \cite{Zagrebnov1,ZagrebnovBook}. In particular,
if $T$ is bounded above with $p$-summable resolvent for a finite $p\geq 1$,
then $T$ is the generator of a Gibbs semigroup, any $T$-bounded perturbation $B$ with relative bound $b<1$
yields a Gibbs semigroup $G_{T+B}$, and the associated Dyson--Phillips expansion \eqref{DDP} holds in the
trace-norm sense and thus \eqref{Trace} holds too \cite[Theorem 4.1]{Zagrebnov1}.

\section{Perturbation theory on $\mathbb{T}^d$}
\label{pert}

Here we develop a perturbation theory for the heat kernel up to the second order
in the potential and in the connection on $\mathbb{T}^d$. On the plane,
corresponding results were obtained long ago \cite{Barvinsky:1987uw,Barvinsky:1990up}.

\subsection{Basic example}\label{sec-be}

The purpose of this subsection is to give the most simple example with which one can
illustrate the technique and discuss global properties of the heat kernel.

Let us take a scalar Laplace operator
\begin{equation*}
L=-\big(\partial_x^2+E(x)\big) 
\end{equation*}
on the unit circle $\T^1$. $E$ is a smooth periodic function, $E\in C^\infty(\T^1)$. According to Section 
\ref{Duhamelexpansion}, the heat trace can be expanded using the Duhamel expansion \eqref{Trace} since it is a 
bounded perturbation of minus the Laplacian.
\begin{equation}
K(L,t)={\rm Tr}\, \big(\exp(-tL)\big) =\sum_{k=0}^\infty K_n(t) \label{Ket}
\end{equation}
giving at the first and second orders in $E$
\begin{eqnarray*}
&&K_1(t)=t\, {\rm Tr}\, \left( e^{t\partial_x^2} E\right) \,,\\
&&K_2(t)=\frac {t^2}2 \,{\rm Tr}\, \left( \int_0^1d\xi \, e^{(1-\xi)t\partial_x^2} \,E\, e^{\xi t\partial_x^2}\, E\right)\,. 
\end{eqnarray*}
After expanding in the Fourier series, one gets
\begin{align*}
K_1(t)&=\frac t{\sqrt{2\pi}} \sum_{q\in \mathbb{Z}}e^{-tq^2} \hat E(0)
=\frac t{2\pi} \sum_{q\in \mathbb{Z}}e^{-tq^2}\int_{\T^1} dx\, E(x)\,,\\
K_2(t)&= \frac{t^2}{4\pi} \int_0^1d\xi \sum_{p,q\in \mathbb{Z}}\hat E(-p)\hat E(p)\, e^{-(q^2+2(1-\xi)pq +(1-\xi)p^2)t}
=\sum_{p\in\mathbb{Z}} \hat E(-p)\,v(p,t)\,\hat E(p), 
\end{align*}
where the form-factor $v(p,t)$ is
\begin{equation}
v(p,t):=\frac{t^2}{4\pi} \int_0^1d\xi \, \sum_{q\in \mathbb{Z}} e^{-(q^2+2(1-\xi)pq +(1-\xi)p^2)t}\label{vp2}.
\end{equation}
This is the analog of $w_\Lambda(p^2)$, $\Lambda^{-2}=t$, in \cite{ILV1}.

The form-factor (\ref{vp2}) may be more explicitly evaluated for small or for large $t$.

Let us first consider the case of small $t$. Physically, this means that the regularization parameter $\Lambda=t^{-1/2}$
is much larger than the inverse radius of $\T^1$ (which we put equal to one in this subsection).
We rewrite
\begin{equation*}
e^{-(q^2+2(1-\xi)pq +(1-\xi)p^2)t}=e^{-((q+(1-\xi)p)^2+\xi(1-\xi)p^2)t}
\end{equation*}
and use the Poisson summation formula to represent
\begin{equation*}
\sum_{q\in \mathbb{Z}}e^{-((q+(1-\xi)p)^2t}=\sqrt{\frac {\pi}t} \sum_{k\in\mathbb{Z}} e^{-2i(1-\xi)p k \pi}\, e^{-k^2\pi^2/t}\,.
\end{equation*}
By dropping the terms that are exponentially small for small $t$ (uniformly in $p$), we arrive at
\begin{equation}
v(p,t)\simeq \frac{t^{3/2}}{4\pi^{1/2}} \int_0^1d\xi\, e^{-p^2\xi(1-\xi)t}\,,\quad t\to 0\,,
\label{vpsmallt}
\end{equation}
which is nothing else than the Barvinsky--Vilkovisky formula \cite{Barvinsky:1987uw,Barvinsky:1990up}, that was
obtained on the plane, but, as we see now, is valid also on $\T^1$ for small $t$.

Similarly, in the first order of $E$ we have for small $t$
\begin{equation*}
K_1(t)\simeq \frac{t^{1/2}}{2\pi^{1/2}} \int dx\, E(x)\,,
\end{equation*}
which is the only term in the heat kernel asymptotics linear in $E$.

The expansion of (\ref{vpsmallt}) for small $p$ reproduces the usual heat kernel expansion, while for large
$p$ we get an analog of the formula obtained in \cite{ILV1}:
\begin{equation}
v(p,t)\simeq \frac 1{2p^2} \sqrt{\frac t{\pi}}\,,\quad t\to 0,\ p\to \infty \,.\label{asv}
\end{equation}
The order of limits here is here with $t\to 0$ first, and then $p\to\infty$.
Below we shall see that the order is not important.

Let us consider the opposite, large-$t$, asymptotic.
 We have
\begin{equation}
K_1(t)\simeq \frac t{2\pi} \int dx E(x) \label{K1larget}
\end{equation}
modulo exponentially small terms. 

To analyse $K_2$,
let us integrate over $\xi$. Suppose, $p\ne 0$.
\begin{equation}
\int_0^1d\xi e^{-t(q^2 +(1-\xi)(2pq+p^2))}=
\frac 1{t(2pq+p^2)} \bigl( e^{-tq^2} -e^{-t(p+q)^2}\bigr) \label{K2lt}
\end{equation}
for $q\ne -p/2$. If $q\ne 0$ and $q\ne -p$, the corresponding contributions
to the sum over $q$ are exponentially small. If $p$ is even, there is a term
with $q=-p/2$ which has to be treated separately. It does not contribute
to the large $t$ asymptotics since the right hand side of (\ref{K2lt})
then equals to $e^{-tp^2/4}$, which is also exponentially small as we assumed
$p\ne 0$. By summing up the contributions from $q=0$ and $q=-p$, one gets
\begin{equation}
v(p,t)\simeq \frac{t}{2\pi p^2}\,,\quad t\to\infty,\ p\ne 0 \label{vplarget}
\end{equation}
up to exponentially small terms. Of course, (\ref{vplarget}) cannot be obtained from (\ref{vpsmallt}).

Another interesting asymptotics of $v(p;t)$ is when $p\to\infty$ at a fixed $t$.
To get this asymptotics, we integrate over $\xi$ in (\ref{vp2}) with the help of
(\ref{K2lt}) and drop all terms with $e^{-tp^2}$, but keep $e^{-tq^2}$
yielding
\begin{equation*}
v(p,t)\simeq \frac t{2\pi} \sum_{q\ne -p/2} \frac{e^{-tq^2}}{(q+p)^2 -q^2} \,,\quad p\to\infty\,.
\end{equation*}
Taking then a $t\to\infty$ asymptotics gives us back (\ref{vplarget}), as expected.
In the limit $t\to 0$ one arrives at (\ref{asv}) where the limits are taken in the reverse order.
This was not guaranteed since the terms $e^{-tp^2}$ are not small in the $t\to 0$ limit.

\subsection{General case}
\label{generalcase}

Consider a Laplace-type operator $L$ acting on sections of a vector bundle
over $\mathbb{T}^d$. 
By choosing an appropriate
covariant derivative $\nabla_\mu =\partial_\mu +\omega_\mu$ and an
endomorphism $E$ one can bring this operator to the form \cite{Gilkey}
\begin{equation}
L=-(g^{\mu\nu}\nabla_\mu\nabla_\nu + E)\,.\label{Lgen}
\end{equation}
We suppose that the metrci $g^{\mu\nu}$ is \emph{constant}. 
The coordinates on $\mathbb{T}^d$
are supposed to be $2\pi$-periodic. Normalization of the Fourier modes
is fixed in a metric-independent way
\begin{equation*}
\phi(x) =(2\pi)^{-d/2} \sum_{k\in\mathbb{Z}^d} \hat \phi (k)\, e^{ikx}\,.
\end{equation*}

We expand the heat kernel as (the notation tr refers to the trace of operators on the vector bundle)
\begin{equation}
K(L,t)=\tr\, \bigl(\sum_{k\in \mathbb{Z}^d} ( \hat E(k) v_1(k,t)+
\hat E(-k) \hat E(k) v_2(k,t) + \hat \omega_\mu (-k) \hat\omega_\nu (k) v_3^{\mu\nu} (k,t)\bigr)
+\dots
\label{v1v2v3}
\end{equation}
where dots denote the terms that are of higher than quadratic order in $E$ and $\omega$.
We also dropped a non-interesting constant term. The form-factors read
\begin{align*}
v_1(k,t)&:=\delta_{k,0} t (2\pi)^{-d/2} \sum_{q\in\mathbb{Z}^d} e^{-tq^2},\\
v_2(k,t)&:=\frac {t^2}{2(2\pi)^d}\int_0^1d\xi \sum_{q\in\mathbb{Z}^d}
e^{-t(\xi (q+k)^2+(1-\xi)q^2)} ,\\
v_3^{\mu\nu}(k,t)
&:=\frac {t^2}{2(2\pi)^d}\int_0^1d\xi \sum_{q\in\mathbb{Z}^d}\bigl[ (k^\mu k^\nu -4q^\mu q^\nu)
e^{-t(\xi (q+k)^2+(1-\xi)q^2)}  +\tfrac 2t g^{\mu\nu} e^{-tq^2} \bigr]\,.
\end{align*}
Here, the vectors with subscripts, $q_\mu, k_\mu$ belong to $\mathbb{Z}^d$, while
$q^\mu\equiv g^{\mu\nu}q_\nu$, etc.

In the small $t$ asymptotic
\begin{equation}
K(L,t)\simeq (4\pi t)^{-d/2}
\int_{\mathbb{T}^d} \sqrt{g} \,dx \,{\rm tr}\, \bigl[ t E + t^2E \tfrac 12 h(-t\partial^2) E
+ t^2 \Omega_{\mu\nu} q(-t\partial^2) \Omega^{\mu\nu} \bigr] +\dots \label{smallt}
\end{equation}
with
\begin{equation}
q(z)\vc -\frac{1}{2} \frac{h(z)-1}{z} \,,\qquad
h(z)\vc\int_0^1 d\alpha \,e^{-\alpha (1-\alpha)\,z}\,. \label{h}
\end{equation}

For large $t$
\begin{equation}
K(L,t)\simeq \frac t{(2\pi)^d}\, {\rm tr}\, \int_{\mathbb{T}^d} dx \bigl[
E(x) + E (-\partial^2)^{-1} E + \tfrac 12 \Omega_{\mu\nu} (-\partial^2)^{-1} \Omega^{\mu\nu} \bigr] +\dots
\label{larget}
\end{equation}

In both (\ref{smallt}) and (\ref{larget}) dots denote higher order terms in $E$ and
$\omega$ which were dropped already in (\ref{v1v2v3}) and also the terms that are
exponentially small in the limits $t\to 0$ and $t \to \infty$ in (\ref{smallt}) and
(\ref{larget}), respectively. By considering different asymptotic regimes in (\ref{v1v2v3})
one arrives at the expressions (\ref{smallt}) and (\ref{larget}) with
$\Omega_{\mu\nu}=\partial_\mu\omega_\nu -\partial_\nu\omega_\nu$.
 However, one can also use the gauge-covariant expression
\begin{equation*}
\Omega_{\mu\nu}=\partial_\mu\omega_\nu -\partial_\nu\omega_\mu + \omega_\mu\omega_\nu-
\omega_\nu\omega_\nu
\end{equation*}
as the difference is in the higher order terms which are neglected anyway.

The following remarks are in order.
\begin{enumerate}
\item Derivation of the formulas above uses the same methods as were employed in
Sec.\ \ref{sec-be} though is considerably more lengthy.
Equation (\ref{v1v2v3})
follows from the Duhamel expansion. To get (\ref{smallt}) on uses the Poisson
summation formula and the drops exponentially small terms. In the opposite
limit, $t\to \infty$, one first integrates over $\xi$ and then neglects exponentially
small contributions.
\item \label{itemii}
As discussed in Section \ref{Duhamelexpansion}, the Duhamel expansion is convergent in the trace-norm. 
No further conditions on $E$ or $\omega_\mu$ (supposed to be $C^\infty (\mathbb{T}^d)$)
are required by (\ref{smallt}) and (\ref{larget}). 
These formulae are good approximations to the heat kernel if
$t^{1/2}$ is much smaller (larger) then the smallest (largest) radius of $\mathbb{T}^n$
defined by the metric $g_{\mu\nu}$. It is interesting to note, that (\ref{smallt})
contains $\sqrt{g}$, while (\ref{larget}) does not. It is important, that the metric is
constant. Although the Duhamel expansion is valid for any smooth metric, it is most useful
when we have a closed expression for the unperturbed heat kernel
$e^{tg^{\mu\nu}\partial_\mu\partial_\nu}$.
\item
The expression (\ref{h}) is the same as obtained for the heat kernel on $\mathbb{R}^d$
\cite{Barvinsky:1987uw,Barvinsky:1990up} where they are valid without the assumption
that $t$ is small. $\mathbb{R}^d$ is, in a sense, an infinite-radii limit of the torus,
so that the small $t$ approximation is always valid (see (\ref{itemii}) above).
\end{enumerate}

\subsection{Consequences for the spectral action}
\label{sec-assa}
We shall not consider here the spectral actions to full generality, but shall restrict ourselves to a particular, 
though rather typical, choice of $f$. Let us take $f=f_e$ with $f_e(z)=e^{-z}$. Then,
\begin{equation*}
\mathcal{S} (\DD,f_e,\Lambda) = K(\DD^2,\Lambda^{-2}) \,,
\end{equation*}
so that most of the results obtained above are valid if one takes $L=\DD_A^2$.
This corres\-ponds to setting
\begin{equation}
\omega_\mu \vc A_\mu,\qquad  E \vc  \tfrac{1}{4}\,[\ga^\mu,\ga^\nu]\,F_{\mu\nu}
\label{oEDD}
\end{equation}
in (\ref{Lgen}). The $\gamma$-matrices are hermitian $m\times m$-matrices with
$m=\left\lfloor d/2 \right\rfloor$ satisfying $\gamma^\mu \gamma^\nu + \gamma^\nu\gamma^\mu =
g^{\mu\nu}$.

Let us consider the large momentum limit of the spectral action. After lengthy but elementary algebra, one obtains
\begin{align}
\mathcal{S} (\DD,f_e,\Lambda) &=-\frac{2^m \Lambda^{-2}}{(2\pi)^d}
\sum_{p,q\in \mathbb{Z}^d} e^{-q^2/\Lambda^2} \,{\rm tr}\, \bigl( 4p^{-6}
p_\rho \hat F^{\rho\mu}(-p)p_\sigma \hat F^{\sigma\nu}(p)q_\mu q_\nu
\nonumber \\ 
& \hspace{5cm}+2p^{-6} p_\rho \hat F^{\mu\nu}(-p)
p_\sigma \hat F_{\mu\nu}(p) q^\rho q^\sigma + O(p^{-6}) \bigr)\label{SAlp}
\end{align}
We see, that the spectral action behaves as $p^{-4}$ at large momenta, as well as on the plane, cf \cite{ILV1}, 
though on the torus the structure of the action is much more complicated. If we now take the limit $\Lambda\to\infty$,
we obtain
\begin{eqnarray*}
&&\mathcal{S} (\DD,f_e,\Lambda)\simeq \sum_{p\in\mathbb{Z}}
{\rm tr}\, [\hat F_{\mu\nu}(-p) w_\Lambda(p^2) F^{\mu\nu}(p)]\,,\\
&&
w_\Lambda (p^2) = - \frac {2^{m+1} \sqrt{g}\Lambda^d}{(4\pi)^m} \, \frac 1{p^4}\,, \quad p\to\infty,\ \Lambda\to\infty\,.
\end{eqnarray*}

We observe the following: (i) the spectral action in this limit has the same universal
behavior $1/p^4$ as was obtained in \cite{ILV1} on $\mathbb{R}^d$; (ii) the same result may be
obtained by taking the $p\to\infty$ limit in (\ref{smallt}) with (\ref{oEDD}) after performing the Fourier transform and 
identification $-\partial^2:=p^2$, i.e., the large $\Lambda$ and large $p$ limits commute.

In the opposite limit, $\Lambda\to 0$, one immediately sees that the whole action (\ref{SAlp}) is exponentially 
small in $\Lambda$, which may also be obtained from (\ref{larget}). In this limit, contributions to the
spectral action from all modes except for the zero mode of $\DD$ are exponentially suppressed.

On the plane, we were able to compute asymptotics of the spectral action for a  fairly general function $f$ \cite{ILV1}. 
The torus case appeared to be much more complicated, but still not hopeless. We shall return to this problem in
a future publication.

\section{Conclusions}
In this paper we considered various expansions of the heat kernel which served
here as a typical example of the spectral action. The standard $t\to 0$ expansion
of the heat kernel (which is local, at least the leading terms) 
is only an asymptotic one and, in general, is not convergent.
On the contrary, the Duhamel (or Dyson--Phillips) expansion (which is non-local)
is convergent with very mild restrictions on the perturbations. Such an expansion is constructed here
for a generic Laplace type operator on $\T^d$ up to the second order in $E$ and $\omega$, giving also
an expansion for the heat kernel of $\DD^2$ up to the second order in $A$. Also discussed are various limiting 
cases, including large/small $t$ (small/large $\Lambda$ is the spectral action), slowly/rapidly varying $A$.

Another meaningful expansion of the spectral action is the one when the curvatures are assumed to be almost 
covariantly constant (that is, derivatives of the curvatures are assumed to be small, though the curvatures 
themselves are of order of unity). Such expansions of the heat kernel were extensively studied by Avramidi \cite{IA}.

\ack
We would like to thank T. Krajewski, T. Sch\"ucker and V. Zagrebnov for fruitful discus\-sions. D. V. V was supported 
by CNPq and FAPESP.

\end{document}